\documentclass[aps,prd,reprint,secnumarabic,nonbibnotes,floatfix,amssymb,longbibliography,showpacs,preprintnumbers]{revtex4-1}

\usepackage{graphicx}
\usepackage{subfigure}
\usepackage{mathptmx}
\usepackage{helvet}
\usepackage{lineno}
\usepackage{textcomp}
\usepackage{multirow}
\usepackage{amsmath}

\begin{document}

\title{ Quantization of the QCD string with a helical structure}

\author{\v{S}\'{a}rka Todorova-Nov\'{a}}  
\affiliation{Institute of Particle and Nuclear Physics, Charles University, Prague}
\email[sarka.todorova@cern.ch]{}

\begin{abstract}   
   The quantum properties of a helix-like shaped QCD string are studied in the context
 of the semi-classical Lund fragmentation model. It is shown how simple quantization
 rules combined with the causality considerations in the string fragmentation process
 describe the mass hierarchy of light pseudoscalar mesons. The
 quantized helix string model predicts observable quantum effects related to the threshold
 behaviour of the intrinsic transverse momentum of hadrons, and of
 the minimal transverse momentum difference between adjacent hadrons.
 The numerical fit of the properties of the QCD field breaking into
 ground state hadrons agree with astonishing precision with values
 obtained recently in the theory of knotted chromoelectric flux tubes.
   
\end{abstract}

\pacs{11.25.-w, 03.75.Lm, 12.40.Yx, 12.38.-t}

\maketitle

\section{Introduction}

  The concept of the QCD string with a helical structure has been introduced
 in \cite{lund_helixm} and some of its potential explored in \cite{helix1}.
 The model has been shown to decribe the experimentally established correlations
 between the longitudinal and transverse momentum components of hadrons measured
 by DELPHI at LEP \cite{z0_delphi} and the azimuthal ordering of hadrons, recently observed by ATLAS at LHC
 (\cite{atlas_ao}).

  The aim of the present paper is to discuss in some detail the space-time 
 evolution of partons following a breakup of a QCD string with a helix
 structure.   A concept of string quantization emerges from these considerations which has the merit to
 describe, in a consistent manner, several experimental observations.   

   The paper is organized as follows: section II describes the transformation of light-cone
 coordinates used by the Lund string model for the case of 3-dimensional string topology
 and introduces the notion of causality in the string fragmentation process.
 Section III deals with the string quantization and the discrete mass spectrum of light mesons.
 Section IV investigates the threshold effect in the transverse
 momentum of direct hadrons. Section V uses the emerging quantization
 model to study the contribution of the intermediate resonant hadronic states to
 the measured particle correlation spectra. In section VI,  the
 similarities between the helix string model and the theory of knotted chromoelectric flux
 tubes are discussed. Section VII closes the paper with a short summary.  
   
\begin{figure}[tbh]
\begin{center}
\includegraphics[width=0.3\textwidth, angle=90]{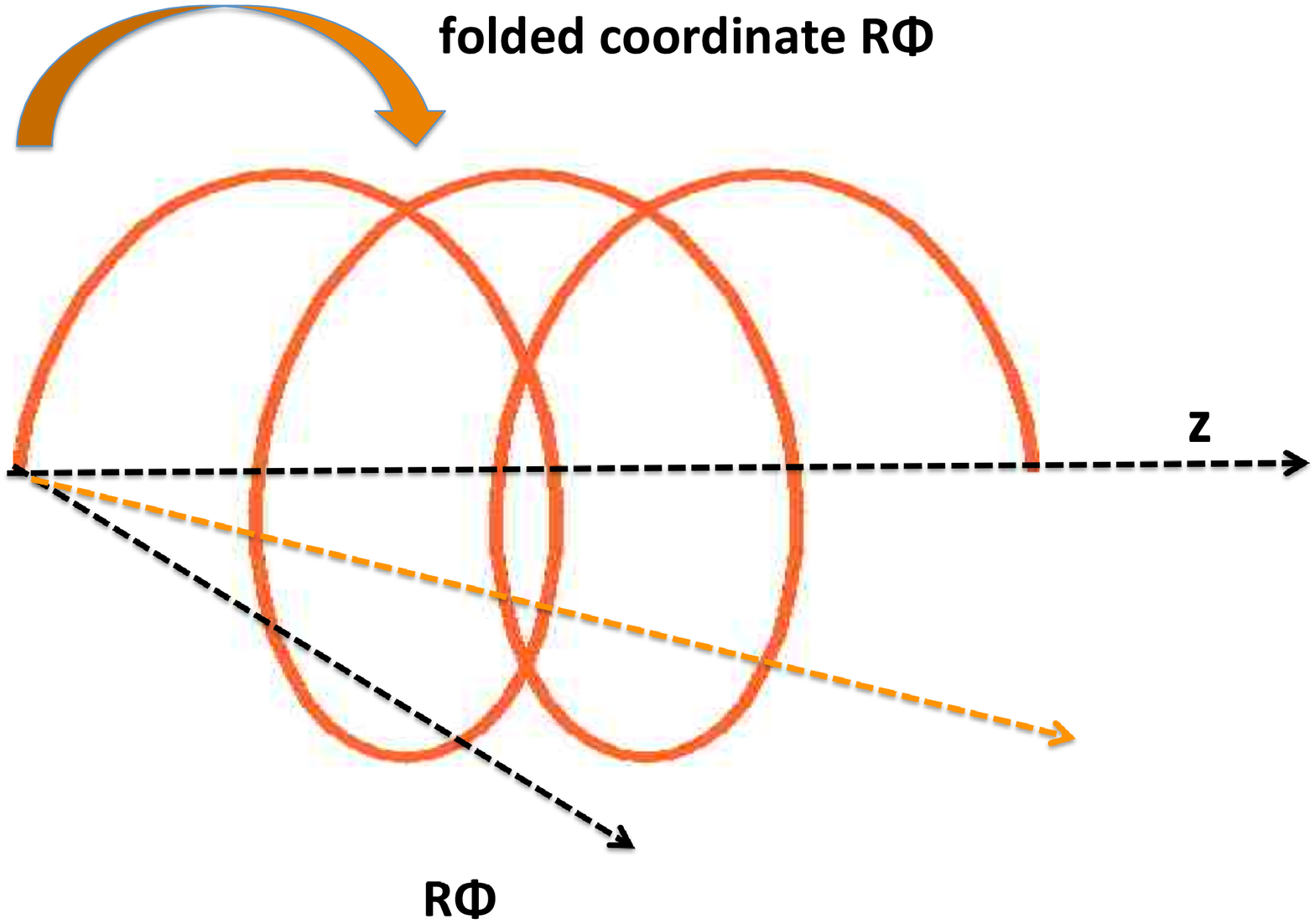}
\caption{ 
 The 2-dimensional coordinate system describing the motion along
 a helix-shaped string consists of the longitudinal string axis $z$  and
 the folded transverse coordinate $R\Phi$, where $R$ stands for radius
 of the helix and the $\Phi$ indicates the helix phase. 
\label{fig:helix}}
\end{center}
\end{figure}

\section{Space-time properties of helical string model}

    In the transition from the 1-dimensional Lund string to a 3-dimensional helix-shaped string,
 it is necessary to reconsider some of the model properties. The basic
 assumption of a  string modelling the confining QCD field with a constant string tension ($\kappa\sim$ 1GeV/fm)
 remains unchanged. However, the use of light-cone coordinates is no longer appropriate,
 as the trajectory of partons in the model is allways bended by the interaction with the field. 
     
   In the case of slowly varying field, the string can be
   approximated by an ideal helix
   with radius $\rm R$ and constant pitch $\rm d\Phi/dz$,  where
   $\Phi$ stands for the azimuthal angle (helix phase) and $z$ is the
   space coordinate parallel to the string axis. Movement of a
   parton along the string can be thus described with the help of the
   longitudinal coordinate $z$ and the folded transverse coordinate
   $\rm R\Phi$ (Fig.~\ref{fig:helix}).
   
    Following a string breakup at $[R e^{i\Phi_B}, z_B, t_B]$ 
  into a pair of massless partons created at rest, the partons will
  move along the string and acquire the momentum 
 \begin{equation}
 \begin{split}
  p_{||}(t) & =   \pm \kappa \beta \ c \ (t - t_B) \\
  p_{T}(t)  & =   \pm \kappa R ( e^{i \omega c (t-t_B)} - e^{i\Phi_{B}} ) 
  \end{split}
  \end{equation}
     
 The longitudinal velocity of partons $\beta$ is related to the
 angular velocity $\omega$ 
  \begin{equation}
  \beta = \sqrt{ 1 - (R\omega)^2 },
  \end{equation}
  ( the light-cone coordinates are recovered in the limit case
  $\rm R\omega = 0$ ) .

   The momentum of a direct hadron created by
 adjacent string breakups at $[R e^{i\Phi_i},z_i,t_i]$, $[R e^{i\Phi_j},z_j,t_j]$ is  
\begin{equation}
\begin{split}
  E_{h} & =  \frac{\kappa}{\beta} | (z_i-z_j) | = \frac{\kappa}{\beta} |\Delta z|,   \\
  p_{h,||} & =   \kappa \beta (t_i - t_j) = \kappa \beta \Delta t, \\
  p_{h,T}  & =   \kappa R ( e^{i\Phi_{i}} - e^{i\Phi_{j}} ),  
\end{split}
  \end{equation}
   and its mass is
  \begin{equation}
   m_{h} = \kappa \sqrt{ (\Delta z/\beta)^2 - (\beta\Delta t)^2 - (2R\sin{\Delta\Phi/2})^2}.  
  \label{eq:mass}
  \end{equation}
  
   There is a fundamental difference (well illustrated by Eq.~\ref{eq:mass})
  between the helical string model and the standard Lund string model
  in what concerns the causality relation 
  between breakup vertices.
 
 In the standard Lund string model, the creation of a massive direct hadron requires a space-like
  distance between breakup vertices ( $\beta=1, m_h>0 \Rightarrow
  |\Delta z|>|\Delta t| $).  The mass spectrum of hadrons is included
  in the model with the help of external parameters.  Arguably,  it is
  the absence of the cross-talk between hadron-generating adjacent
  string breakups which prevents the model from developing physics
  scenarios investigating the origin of the hadron mass
  hierarchy.   

   The situation is different in the case of the helical string model,
   where a time-like distance between breakup vertices is possible. 
   As shown below, the causality considerations seem to pave the way to
   a better understanding of the role of quantum effects in the
   fragmentation.  In order to explore the causal properties of the model,
   the time-like distance between adjacent breakup
   vertices will be imposed and the consequences studied. ( This does not
   necessarily mean the  space-like separation between breakup
   vertices should not occur in the fragmentation - see the discussion in Section VI. )  
    
   There is an ambiguity concerning the way the signal is allowed to propagate.
   If the information  ( about a breakup of the string at a given
   point) is allowed to pass along the string only, the space-time distance
  between adjacent vertices becomes negligible (to the extent we have neglected the parton masses)
  which means the propagating parton essentially triggers the following break-up and the mass
  of the outcoming hadron is (note that in this case $\Delta z = \beta \Delta t$ )
  \begin{equation}
   m_S (\Delta\Phi) = \kappa R \sqrt{ (\Delta\Phi)^2 - (2\sin{\Delta\Phi/2})^2 }. 
  \label{eq:mass_s}
  \end{equation}  
   
    It is interesting to see that the longitudinal momentum is factorized out from the equation
  and that the hadron mass depends on the transverse properties of the string shape only.
  To obtain a discrete mass spectrum, it is sufficient to introduce quantization of
  the transverse coordinate $\rm R\Phi$ (to be discussed in the following section).  

\begin{figure}[h!]
\begin{center}
\includegraphics[width=0.25\textwidth, angle=90]{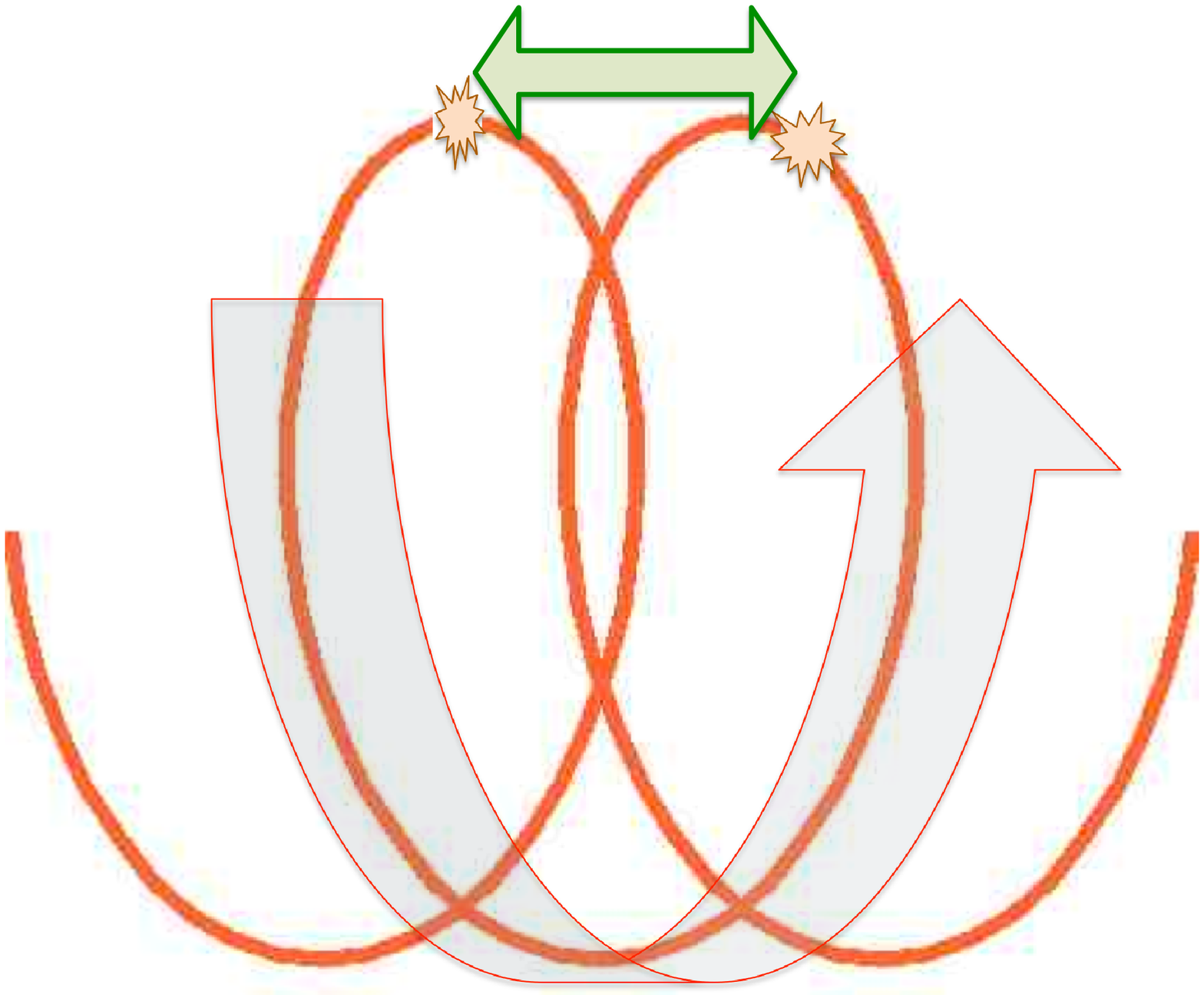}
\caption{ 
 The information about the string breakup propagates preferably along the string field,
 though a cross-talk between string loops is not excluded, either.
 The requirement of causal relation between breakups leads to an
 effective decoupling of the longitudinal and transverse component of
 the hadron momenta in the former case (see text).
\label{fig:vortex}}
\end{center}
\end{figure}

    There is of course also a possibility that the information about the breakup travels inside the string vortex (Fig.~\ref{fig:vortex}).
  To maintain the time-like difference between string breakups in such
  a case, the allowed time difference is then restricted to the
  interval 
 \begin{equation} 
      \sqrt{ (\Delta z)^2+(2R\sin{\Delta\Phi/2})^2 } \leq c \Delta t \leq \sqrt{ (\Delta z)^2+(R\Delta\Phi)^2 }
\end{equation}
    and the outcoming hadron has a mass $m_C$ in the range
 \begin{equation}
        m_S (\Delta\Phi)   \leq m_C (\Delta\Phi)   \leq  m_S (\Delta\Phi)  \sqrt{1 + \beta^2} 
 \label{eq:mass_c}
\end{equation}
 (the subscripts $\rm S,C $ stand for ``singular'' and ``continuous''
 mass solutions).

   The transverse energy $E_T$ of the hadron, which coincides with the
   transverse mass $m_T = \sqrt{ m^2 + p_T^2 }$ in the ``singular'' case,
 does not depend on the choice (or the abandon) of the causality scheme
 and follows the simple relation
\begin{equation}
     E_{T}(\Delta\Phi)  =  \kappa R\Delta\Phi.
\end{equation}

\section{Mass spectra }
 
    Building on the causality requirements, we have obtained relations between the transverse string properties
  and the allowed hadron mass spectrum. It seems only natural to take a step further and to try to establish
  a quantization pattern for the string fragmentation which would match the measured discrete hadron mass spectra.
  
   Let's assume the string quantization is realized through the quantization of the transverse coordinate
 \begin{equation}
    R\Phi \Rightarrow  \rm{n} R\Delta\Phi = \rm{n} \xi,  ( n=1,2,.. )   \\
 \end{equation}  
       and that the n=1 case corresponds to the lightest hadron,  the $\pi$ meson.

   Eq.~\ref{eq:mass_s} is particularly interesting for the study of light meson mass hierarchy because it describes
  the narrow pseudoscalar states (PS) decaying into an odd number of pions 
 \begin{equation}
   \begin{split}
   PS  & \rightarrow  \rm{n} \pi,  \ \rm{n}=(1),3,5,..  \\
   m(PS) & =  \kappa \sqrt{ (\rm{n}\xi)^2 - (2\xi/\Delta\Phi)^2 sin^2(\rm{n}\Delta\Phi/2)}.
   \end{split}
 \label{eq:mass_PS}
 \end{equation}
   
 The results of the best fit matching the
 Eq.~\ref{eq:mass_PS} to experimentally measured data \cite{pdg} are
 listed in Table~1. Despite the fact that the
 simultaneous fit of 2 unknowns ($R,\Delta\Phi$) from 3 hadronic
 states is overconstrained,  a common solution describing the
 properties of the ground state is found.   The $\pi$, $\eta(548)$ and $\eta'(958)$
 masses are reproduced by Eq.~\ref{eq:mass_PS} with precision better than 3\% using $\xi = 0.192$ fm and $\Delta\Phi$=2.8
 (for $\kappa=$1 GeV/fm).  

\begin{table}[t!]{
\begin{center}
\begin{tabular}{|c|c|c|}
\hline
 $\kappa \xi$ [MeV] &  $\kappa$ R [MeV]  &  $\Delta\Phi$  \\
\hline
    192.5 $\pm$ 0.5   &       68 $\pm$ 2     & 2.82 $\pm$ 0.06  \\
\hline \hline
  meson   &     PDG mass [MeV]    &   model estimate [MeV] \\
\hline
   $\pi$    &     135 - 140     &    137  \\
   $\eta$   &       548         &    565   \\        
   $\eta'$  &       958          &     958   \\
\hline
\end{tabular}
\label{tab:meson}
\caption{ Best fit of the parameters of the pion ground state obtained
  from the mass spectrum of light pseudoscalar mesons. The $\eta$ mass
  is reproduced within a 3\% margin which serves as the base of 
  uncertainty for $R,\Delta\Phi$ parameters.}
\end{center}}
\end{table}      

\begin{figure}[bh!]
\begin{center}
\includegraphics[width=0.45\textwidth]{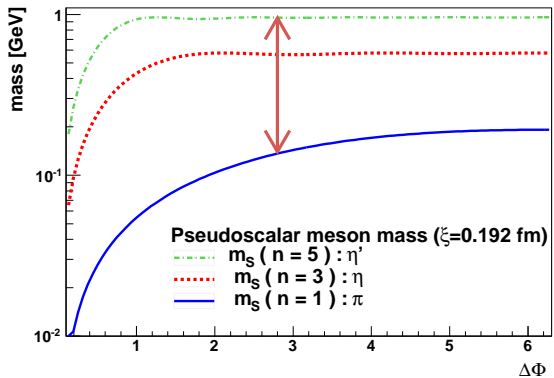}
\caption{ 
  The predicted masses of light pseudoscalar mesons as function of helix phase difference $\Delta\Phi$,
 for fixed $R\Delta\Phi$=0.192 fm rad.
\label{fig:PS}}
\end{center}
\end{figure}

 Fig.~\ref{fig:PS} shows the dependence of the mass of PS mesons as a function
 of $\Delta\Phi$ in Eq.(~\ref{eq:mass_PS}). With increasing $\Delta\Phi$, the predicted masses of $\eta$ and $\eta'$ reach the plateau
 (around $\Delta\Phi \sim 1.5$ rad) and lose sensitivity to the
 $\Delta\Phi$ value, but the mass of the $\pi$ meson
 rises steadily till  $\Delta\Phi \sim 5$ rad and effectively fixes the $\Delta \Phi$ value in the model.    

   The scalar nature of PS states is in agreement with the expectations of the quantization model:
  $m_{T}(PS)=\rm{n} \ \it{m}_{T}(\pi)$, thus the decay products
  of ($\eta,\eta'$) have negligible (longitudinal) relative momentum in the rest frame of the mother resonance.  
 
   If the quantization model, in the first approximation, fits the mass spectra of light PS mesons,
 what can be said about the vector mesons (VM) ?
  
   The lightest vector mesons $\rho(770)$ and $\omega(782)$ can be interpreted as n=4 states decaying
 into m $<$ n pions:
 \begin{equation*}
     m_{S}(n=4)=0.76 \ \rm{GeV},
\end{equation*}
  or n=3 states formed according to Eq.~\ref{eq:mass_c}:
\begin{equation*}
 \sqrt{2} \ m_S(n=3) \sim 0.79 \ \rm{GeV},
\end{equation*}
    and their non-zero total angular momentum arises from the relative momentum of decay products
 (kinematically allowed since  $m_{T}(VM) > \rm{m} \ \it{m}_{T}(\pi)$.

 The mass of K$^*(890)$ and $\Phi(1020)$ mesons 
 can be roughly associated with the mass of the $\rho(770)$ increased
 by the mass of the strange quark(s) ($\sim$120 MeV).
 (The same reasoning would classify K meson as a $n=2$ state). 

  The tentative classification of VM hadron states mentioned above
 is more an illustration than a prediction ( a more precise
 relation between the total angular momentum of the hadron and the
 field properties has yet to be established ).  It is worth noticing
 however that the quantization of the transverse component of the
 string is equivalent to the quantization of the angular momentum $J$
 stored in the string ( proportional to the transverse area spanned by
 the string ) and that the relation 

\begin{equation}
J \simeq \kappa (R \Delta\Phi)^2  =  m_{T}^{2} / \kappa      
\end{equation}
     
  indicates that the spectra derived from the model will lie along 
  Regge trajectories \cite{regge}.

\section{Transverse momentum threshold}

\begin{figure}[h!]
\begin{center}
\includegraphics[width=0.4\textwidth]{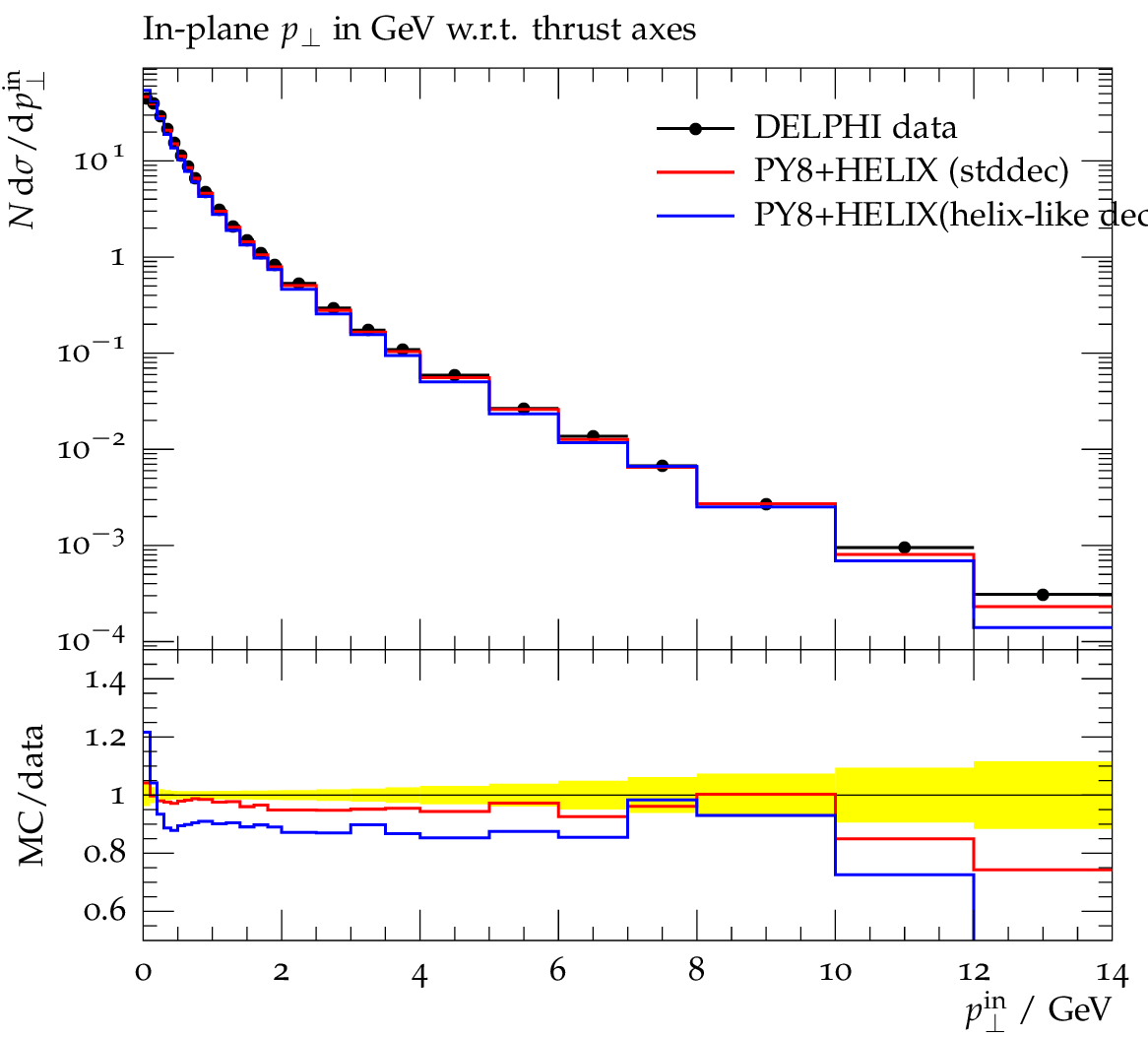}
\includegraphics[width=0.4\textwidth]{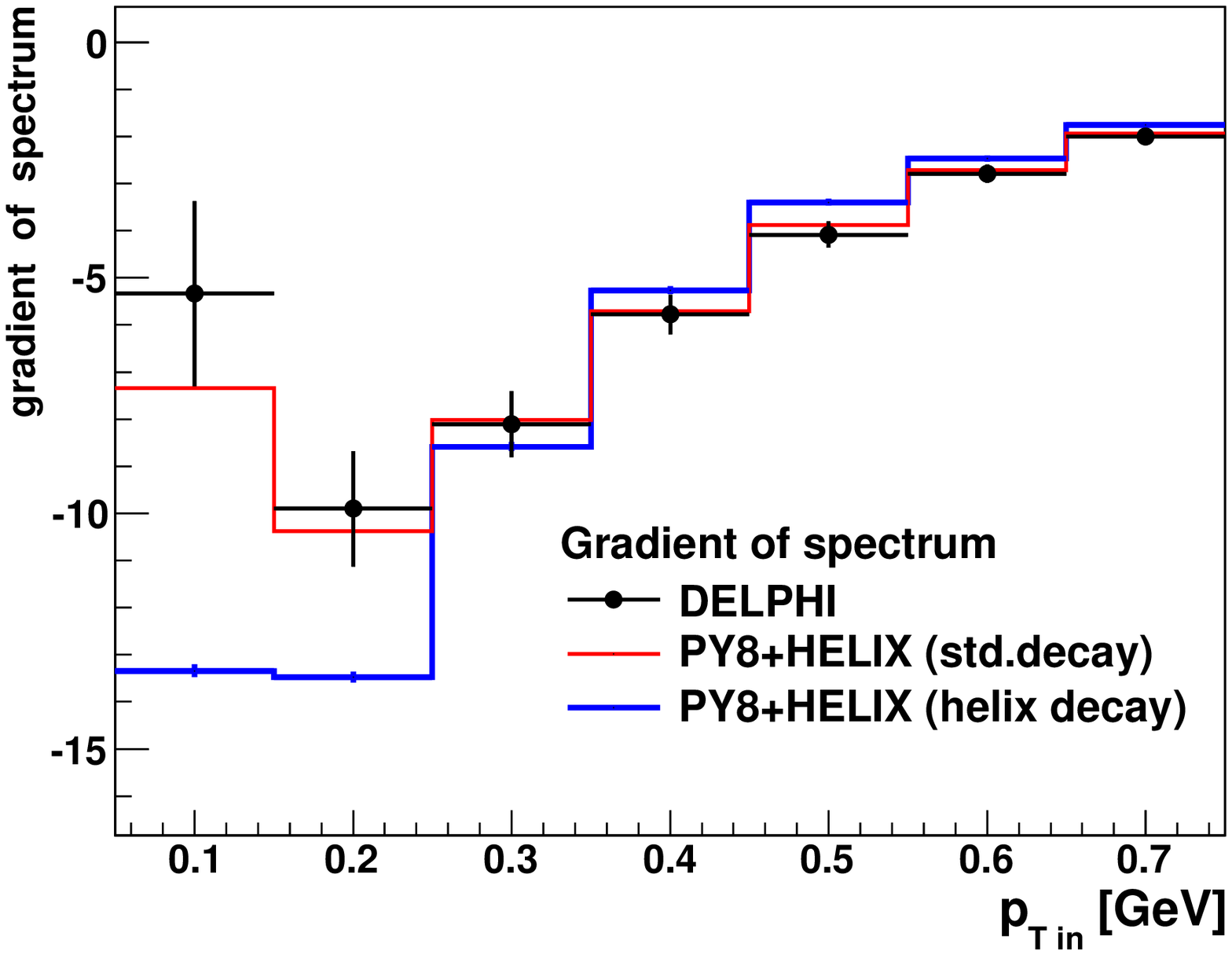}
\caption{ 
 Comparison of the helix string model predictions with the DELPHI data \cite{z0_delphi}.
 Upper plot: The production of low $p_T$ charged particles is overestimated by the
 model when decay of short lived resonances is treated as a smooth
 continuation of the fragmentation of the helix string.
 Lower plot : Comparison of gradients of the inclusive distributions. The extended
 helix string model overestimates the rate of particles with low
 $p_T$.   The most significant change of slope occurs in the data
 below $p_T \sim$ 0.2 GeV - this may be related to the
 existence of an intrinsic pT cutoff. 
\label{fig:pt}}
\end{center}
\end{figure}

  The discretization of the mass spectrum is not the only quantum
  effect which can be observed in the string fragmentation. In fact,
  the current investigation of the properties of the helix string
  quantization was prompted by the study of the inclusive $p_T$ spectra.
  In \cite{helix1} it has been shown that the helix string model
  significantly improves the description of the inclusive low $p_T$
  region. It has been also shown that the strength of the
  azimuthal correlations between hadrons  can be described by the model
  only if the helix string model is extended to the decay of
  short-lived resonances. However, it turns out that such an extension
  spoils the agreement between the LEP data and the helix model
  essentially because the resonance decay according to the helix shaped
  ``field memory'' produces way too many low $p_T$ particles
  (Fig.~\ref{fig:pt}). 

The effect
  cannot be tuned away as there are essentially no relevant free parameters
  left in the model.  A careful study of the discrepancy and the
  gradients of the $p_T$ spectra does not exclude existence of a
  natural $p_T$ cutoff just below 0.2 GeV (Fig.~\ref{fig:pt}, bottom plot).

     It is therefore encouraging to see - on the basis of results obtained in
  the previous section - that the production of soft pions with the $p\sim p_{T}  < $
  0.14 GeV should be supressed in the quantized model. This result has yet to be propagated through the entire fragmentation and decay
chain but this particular model feature is expected  to help the regularization
of the soft particle production in
the helix string model extended to the decay of resonances.

\section{Momentum difference of adjacent hadrons}

   The quantization of the helix string implies a quantization of the
   momentum difference between adjacent hadrons.  Since 
   the local charge conservation forbids the production
   of adjacent like-sign charged hadron pairs in the fragmentation process,     
   the quantum effects can play a large role in the correlation
   phenomena with a significant
   difference between particle pairs with like-sign and unlike-sign charge
   combination. 
  
\begin{figure}[tbh]
\begin{center}
\includegraphics[width=0.5\textwidth]{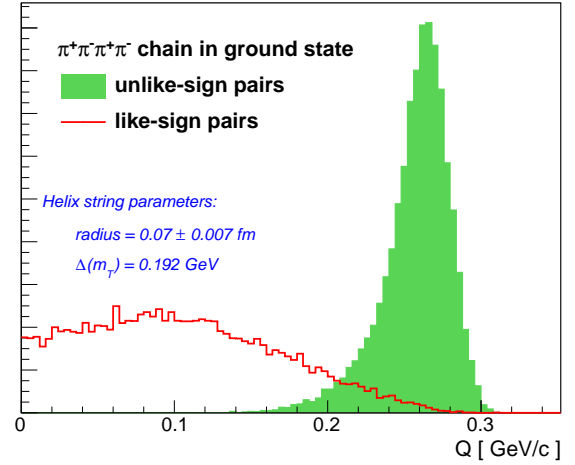}
\caption{ The correlation pattern estimate for a chain of 4 charged
  pions in the ground state (Table 1). The longitudinal momentum
  differences are neglected, a variation of 10\% is applied on the
  helix radius instead, in order to obtain a smooth spectra.
\label{fig:q_chain}}
\end{center}
\end{figure}

     In the approximation of an ideal, or slowly varying helix
    string field, it is possible to make an estimate of the charge
    combination asymmetry induced by quantization. Consider
    a chain of adjacent charged pions in the ground state, for example
    from $\eta'$ decay. The homogenity of the QCD field
    implies the difference between longitudinal momenta components 
    of such pions is negligible. The momentum difference between
    pions along the chain is then given by the helix phase difference

 \begin{equation}
\begin{split}
     Q & = \sqrt{-(p_i - p_j)^2} \approx 2 p_T  | \rm sin[ 0.5( \Phi_i -  \Phi_j)] | \\
    &  =  2 p_T | \rm sin [ 0.5 ( j-i ) \Delta\Phi ] |, 
\end{split}
  \end{equation}  

  where $p_T (\sim 0.14 $ GeV) is the transverse momentum of the pion
  in the ground state, $i,j = 0,1,2,3$ are integers corresponding to the rank of
  the hadron along the chain, and $\Delta\Phi (\sim 2.8 )$ is the
  opening azimuthal angle between adjacent ground state pions.
  Fig.~\ref{fig:q_chain} shows the resulting correlation pattern with
  a marked separation of like-sign and unlike-sign pairs (the helix
  radius has been randomly varied by 10\% in order to produce a smooth
  spectrum). The onset of excess of like-sign pairs occurs at $ Q\sim
  0.2 $ GeV in the model. A large amount of experimental data provides
  evidence of an excess of like-sign hadron pair production in
  the low Q region. Most often, the data are studied from the
  perspective of the Handbury-Brown-Twiss model, i.e. as a signature
  of the incoherent particle production. The helix string model
  suggests an alternative point of view - such correlations
  may well be associated with fully coherent hadron
  production.  In the specific case under study,  due to the large
  opening angle $\Delta\Phi$, the quantized chain of ground state pions 
  acquires properties reminiscent of Bose-Einstein condensate.

   It should be possible to make a more precise experimental
   evaluation of the role of hadron 'chains' (and $\eta'$ decay) in the correlation
   signal. Such a study may have a significant impact on the further
   development of the helix model, as it may confirm, or
   reject, the hypothesis of a strong link between resonance production
   and correlation phenomena. 

\section{Discussion of results}

   The radius of the quantized helix string obtained in Section III is significantly
   smaller than the tuned radius of the non-quantized helix
   string and will require further adjustments in the fragmentation
   model in order to describe the data ( via radial string
   excitations, gradual field relaxation, or adjustment of the parton
   shower cutoff ).  On the other hand, it is fair to admit that, though somewhat
   surprising, the numerical values describing the ground hadronic
   state do provide a basis for explanation of the charged combination
   asymmetry in 2-particle correlations, and to the regularization of
   the soft particle production, without introduction of additional
   free parameters in the model.

    Moreover, the estimated
    effective radial size of the helix string breaking into a set of
    ground state hadrons ( $ 68 \pm 2$ MeV ) turns out to be
    in agreement with the estimate of the confinment scale
    $ \Lambda_{tube} = 65.16 \pm 0.61 $ MeV   obtained from the fit of $J^{++}$ spectra
    modelled via tight topological QCD knots  \cite{buniy&kephart}.
          
     The agreement is remarkable not only because the two studies
     use a different type of input data ( mesons/open strings versus
     glueballs/closed strings ) but also due to the fact that reasoning behind
     the two semi-classical models follows distinct paths, one
     \cite{lund_helixm} being based on study of the gluon emission,
     the other \cite{buniy&kephart} stemming from the physics of
     continuum (plasma) and topological properties of QCD flux tubes. In
     both models, the helicity conservation plays a central role, and
     the approach to the problematics seems to be fairly consistent up
     to the point that a certain aspects of one model can be
     intuitively  understood in terms of the alternative approach.
     For example, the secondary emission of gluons in the helix string
     model, which leads to the homogenization of the string
     field (prefered by the data), can possibly be seen as the ``field relaxation into equilibrium state with
     minimal energy''  - the term employed by the model of knotted QCD
     flux tubes.  

      As a final remarque, it should be emphasized the helix string
      model is expected to remain conformant to the string area law
      governing the string breakup probability. That means the model
      should be able to handle the fragmentation resulting from
      several independent ( uncorrelated, space-like ) string breakups
      followed by a chain of correlated breakups.

\section{Conclusions}

     The properties of the quantized helix string model have been
     investigated using a data driven simple quantization recipe.
     The model allows to introduce proper causal relations between 
    the breakup vertices in the string fragmentation. The causality
     represents a strong constraint for the particle production and
     helps to understand the emergence of narrow hadronic resonant
     states. The fit of the light pseudoscalar mesons provides 
     the parameters describing the ground hadronic state, and allows
     to make predictions concerning the threshold behaviour of relevant
     observables, to be verified with the help of experimental data.
     Last but not least,  the fitted estimate of the helix string
     radius ($\kappa$R = 68+-2 MeV)  is found
     to be in good agreement with the results of a recent study
     \cite{buniy&kephart} of mass spectrum of  $J^{++}$ states
     (glueball candidates) in the frame of the model of knotted QCD flux tubes.

\section{Acknowledments}

   The author would like to thank Prof.J.Bj{\o}rken for valuable
   suggestions.

\bibliography{draft}

\end{document}